\documentclass[usenatbib]{mn2e}

\usepackage{epsfig}
\usepackage{mathrsfs,aas_macros}

\def\Lya{Ly$\alpha$~}

\def\HI{\hbox{H~$\rm \scriptstyle I\ $}}

\def\HeII{\hbox{He~$\rm \scriptstyle II\ $}}

\title[The flux distribution of the \Lya forest]{Possible evidence for
an inverted temperature-density relation in the intergalactic medium
from the flux distribution of the \Lya forest}

\author[J.S. Bolton et al.] {J.S.
  Bolton$^{1}$, M. Viel$^{2,3}$, T.-S. Kim$^{4}$, M.G. Haehnelt$^{5}$ \& R.F. Carswell$^{5}$ \\
  $^1$ Max Planck Institut f{\"u}r Astrophysik, Karl-Schwarzschild
  Str. 1, 85748 Garching, Germany \\
  $^2$ INAF-Osservatorio Astronomico di Trieste, Via G. B. Tiepolo 11,
  I-34131 Trieste, Italy\\
  $^3$ INFN/National Institute for Nuclear Physics, Via Valerio 2,
  I-34127 Trieste, Italy\\
  $^4$ Astrophysikalisches Institut Potsdam, An der Sternwarte 16,
  D-14482 Potsdam, Germany\\
  $^5$ Institute of Astronomy, University of Cambridge, Madingley
  Road, Cambridge,  CB3 0HA \\}

\begin{document}

\date{20 February 2008}

\maketitle

\label{firstpage}

\begin{abstract}
We compare the improved measurement of the \Lya forest flux
probability distribution at $1.7<z<3.2$ presented by Kim et al. (2007)
to a large set of hydrodynamical simulations of the \Lya forest with
different cosmological parameters and thermal histories. The
simulations are in  good agreement with the observational data if the
temperature-density  relation for the low density intergalactic medium
(IGM), $T=T_0 \Delta^{\gamma-1}$, is either close to isothermal or
inverted ($\gamma<1$). Our results suggest that the voids in the IGM may be
significantly hotter and the thermal state of the low density IGM  may
be substantially more complex than is usually assumed at these
redshifts.   We discuss radiative transfer effects which alter  the
spectral shape of ionising radiation during the epoch of \HeII
reionisation as a possible physical mechanism for achieving an
inverted temperature-density relation at $z\simeq 3$.

\end{abstract}
 

\section{Introduction}

Traditional analyses of the \Lya forest observed in the spectra of
high redshift quasi-stellar objects (QSOs) decompose the flux
distribution into a series of discrete absorption profiles, generally
characterised by a Voigt function
(\citealt{Hu95,Lu96,KirkmanTytler97,Kim97,Kim01}).  Voigt profiles
provide an accurate  description for an absorption line if the
absorber is a localised  gas cloud with a Gaussian velocity dispersion
(\citealt{Rauch98}). In the now widely established paradigm for the
origin of the \Lya forest most of the absorption is, however, caused
by extended gas distributions broadened by the Hubble flow, removing
much of the physical motivation for the decomposition of the flux
distribution into Voigt profiles. Such a decomposition is nevertheless
useful, but it is non-unique and the line fitting process is time
consuming and somewhat subjective (see \citealt{Kim07}, hereafter K07,
for an exhaustive discussion).  Alternative characterisations of the
flux distribution based on pixel statistics (\citealt{Jenkins91,
Rauch97,Gaztanaga99,McDonald00,Theuns00,Meiksin01,Viel04c,Lidz06b,Becker07})
and wavelet analysis
(\citealt{Meiksin00,TheunsZaroubi00,Zaldarriaga02}) have also been
developed. These measures are simple and - at least in principle -
easy to compare to the same quantities extracted from theoretical
models of the \Lya forest
(\citealt{Cen92,MiraldaEscude96,Bi97,Croft99, Dave99,Theuns98,
Jena05}).  In particular, the power spectrum of the \Lya forest flux
distribution has recently been successfully developed into a
quantitative tool for measuring the matter power spectrum on scales of
$0.5 h^{-1}$ Mpc to $40h^{-1}$ Mpc
(e.g. \citealt{Croft02,Viel04b,McDonald05}).

The simplest pixel statistic is the \Lya flux probability distribution
function (PDF), which is sensitive to the density distribution and
thermal state of the IGM ({\it e.g.} \citealt{Becker07}).  Several
attempts have been made to obtain joint constraints on a variety of
cosmological and astrophysical parameters using the flux PDF together
with the flux power spectrum (\citealt{Choudhury01,
Desjacques05,Lidz06b,Desjacques07}).  However, the PDF is also
sensitive to a range of systematic uncertainties, most notably the
assumed continuum level, the noise properties of the spectra and
contamination by metal absorption lines in the \Lya forest.  In high
resolution, high signal to noise data at  $z<3$, the first two
uncertainties have a small impact on the shape of the PDF (K07). The
$z<3$ metal absorption in the \Lya forest, however, alters the shape
and amplitude  of the PDF significantly, especially at
$0.2<F<0.8$. Since many numerical simulations of the \Lya forest do
not include metals, and those that do must also correctly reproduce
their uncertain distribution and abundance, this can lead to
substantial systematic uncertainties when attempting to fit simulation
data to the observed flux PDF of the \Lya forest.

K07 have completed a detailed analysis of the impact of metal
absorption lines on the \Lya PDF at $1.7<z<3.2$.  They presented new
measurements of the PDF in which all identifiable metal contamination
has been removed from the \Lya forest on a line-by-line basis.  The
metal cleaned PDF had a significantly different shape when compared
directly to the metal contaminated sample.  In this paper we shall
investigate what can be inferred about the underlying matter
distribution and the thermal state of the intergalactic medium from
these new PDF measurements using state-of-the-art hydrodynamical
simulations of the \Lya forest.  Our approach is different
to most recent studies of the flux PDF which have used either
pseudo-hydrodynamical techniques, dark matter only simulations or
semi-analytical models
(\citealt{Choudhury01,Desjacques05,Lidz06b,Becker07}) which model 
the effect of gas pressure on the underlying IGM density distribution
in an approximate way.

The structure of this paper is as follows. We discuss the
observational data and the hydrodynamical simulations in section 2.
In section 3, we qualitatively compare the simulation data to the
observations and determine the impact of various astrophysical and
numerical effects on the simulated flux PDF.  A quantitative
comparison is undertaken in section 4 and the best fitting models to
the K07 data are presented along with a comparison to other recent
studies in the literature.  We briefly discuss the implications of our
results for measurements of the matter power spectrum from the \Lya forest
flux power spectrum in section 5 and conclude in section 6.

\section{Observations and simulations of the \Lya forest}
\subsection{The observational data}

The \Lya flux PDF at $1.7<z<3.2$ was recently measured by K07 using a
set of 18 high resolution ($R \sim 45~000$), high signal to noise (S/N
$\geq 30-50$) VLT/UVES spectra.  Metal lines were removed from the
\Lya forest by directly fitting all identifiable absorption features.
This approach to metal absorption removal is different to other
methods in the literature, which either account for metal
contamination statistically (\citealt{Tytler04,Kirkman05}) or by
excising heavily contaminated regions from the spectra
(\citealt{Rauch97,McDonald00,Lidz06b}). Following this procedure, K07
obtained the \Lya flux PDF in three redshift bins at $\langle z
\rangle =2.07$, $\langle z \rangle =2.52$ and $\langle z \rangle
=2.94$.  We shall compare these new improved measurements to the PDF
computed by numerical simulations of the  \Lya forest. Further details
regarding the observational data and its reduction may be found in K07.

\subsection{Numerical simulations of the IGM}

The simulations used in this study were run using the parallel
Tree-SPH code $\rm \scriptstyle GADGET-2$ (\citealt{Springel05}),
modified to compute the ionisation state of the gas using the
non-equilibrium ionisation algorithm of \cite{Bolton06}.  The fiducial
simulation volume is a periodic box $20h^{-1}$ comoving Mpc in length
containing $2 \times 256^{3}$ gas and dark matter particles.   The
gravitational softening length was set to $1/30^{\rm th}$ of the mean
linear interparticle spacing.  The gas in the simulations is
photoionised by a spatially uniform ultraviolet background (UVB) based
on the galaxies and QSOs emission model of \cite{HaardtMadau01}.  The
mean free path for hydrogen ionising photons is expected to be much
larger than the mean separation between ionising sources at $z<3$, and
so the assumption of a uniform UVB at the hydrogen Lyman limit should
be a reasonable one (\citealt{BoltonHaehnelt07b}).  Star formation is
included using a simplified prescription which converts all gas
particles with overdensity $\Delta > 10^{3}$ and temperature
$T<10^{5}\rm~K$ into collisionless stars, significantly speeding up
the simulations.  The runs were all started at $z=99$, with initial
conditions generated using the transfer function of
\cite{EisensteinHu99}.  The simulations explore a range of
cosmological and astrophysical parameters consistent with other
observational data, and are summarised fully in Table$~1$.

\begin{table*}
\centering
\caption{Parameters used in our suite of hydrodynamical simulations.
In all models we assume a flat Universe with
$\Omega_{\Lambda}=1-\Omega_{\rm m}$, $\Omega_{\rm m}=0.26$,
$\Omega_{\rm b}h^{2}=0.024$, $h=0.72$ and $n=0.95$.  The
temperature-density relation is estimated at each redshift by a least
squares fit to the volume weighted temperature-density plane at $0.1
\leq \Delta \leq 1$ and $T_0 \leq 10^{5} \rm~K$.  The $T_0$ values
have been rounded to three significant figures.}
\begin{tabular}
{c|c|c|c|c|c|c|c|c|c|c}
  \hline
  Model & $\sigma_{8}$ & $T_{0}$ [K] & $T_{0}$ [K] & $T_{0}$ [K]
  &$\gamma$ & $\gamma$ & $\gamma$ & $\alpha$ & $\beta$ & Notes  \\ 

   & & ($z=2.07$) & ($z=2.52$)  & ($z=2.94$) & ($z=2.07$)  &
  ($z=2.52$) & ($z=2.94$) &  & & \\
  \hline
 20-256   & 0.85 & 14800 & 17600 & 20800 & 1.37 & 1.31 & 1.27 & 1.0 & 0.0
  & Fiducial model\\

 20-100   & 0.85 & 16800 & 19600 & 22700 & 1.40 & 1.35 & 1.30 & 1.0 & 0.0
  & Low resolution \\
 20-200   & 0.85 & 15100 & 18000 & 21300 & 1.38 & 1.33 & 1.29 & 1.0 & 0.0
  & Med. resolution \\
 20-400   & 0.85 & 14400 & 17200 & 20300 & 1.37 & 1.31 & 1.26 & 1.0 & 0.0
  & High resolution. \\

 40-200   & 0.85 & 16900 & 19700 & 22700 & 1.42 & 1.36 & 1.31 & 1.0 & 0.0
  & Low res./Med. box \\
 80-400   & 0.85 & 16900 & 19800 & 22800 & 1.41 & 1.36 & 1.31 & 1.0 & 0.0
  & Low res./Large box \\

 20-256m   & 0.85 & 14800 & 17700 & 20900 & 1.37 & 1.31 & 1.27 & 1.0 & 0.0
  & Multiphase SF\\
 20-256w   & 0.85 & 18900 & 20900 & 23300 & 1.47 & 1.39 & 1.32 & 1.0 & 0.0
  & Strong winds\\

 20-256t1  & 0.85 & 10900 & 12600 & 14600 & 1.41 & 1.36 & 1.31 & 0.5 & 0.0
  & Low $T_{0}$ \\ 
 20-256t2   & 0.85 & 17600 & 21300 & 25500 & 1.35 & 1.29 & 1.25 & 1.4 & 0.0
  & High $T_{0}$ \\ 

 20-256g1   & 0.85 & 15300 & 18300 & 21700 & 1.59 & 1.56 & 1.53 & 1.0 & 0.5
  & High $\gamma$ \\
 20-256g2   & 0.85 & 14100 & 16700 & 19800 & 1.16 & 1.08 & 1.02 & 1.0 &
  -0.4 & Low $\gamma$ \\
 20-256g3   & 0.85 & 12600 & 14700 & 17200 & 0.67 & 0.54 & 0.44 & 1.0 &
  -1.2 & Inverted $\gamma$ \\
 20-256xs   & 0.85 & 15900 & 18800 & 21800 & 1.38 & 1.32 & 1.27 & 1.0 &
  0.0 & Extra scatter \\

 20-256s1   & 0.75 & 14600 & 17400 & 20700 & 1.37 & 1.32 & 1.27 & 1.0 & 0.0
  & $\sigma_8=0.75$\\
 20-256s2   & 0.95 & 15000 & 17700 & 21000 & 1.37 & 1.31 & 1.26 & 1.0 & 0.0
  & $\sigma_8=0.95$ \\

 20-256s2g2 & 0.95 & 14200 & 16700 & 19700 & 1.15 & 1.07 & 1.01 & 1.0 & -0.4   & $\sigma_8=0.95$, Low $\gamma$ \\

 \hline
\end{tabular}
\end{table*}

Constraints on the thermal state of the low density IGM are usually
expressed in terms of a polytropic temperature-density
relation\footnote{The temperature-density relation is also sometimes
referred to as the IGM effective equation of state, or simply just the
equation of state.  Note, however, that the equation of state for this
low density gas in the correct sense is the ideal gas equation.},
$T=T_{0}\Delta^{\gamma-1}$, expected to arise for $\Delta = \rho /
\langle \rho \rangle \leq 10$ when photo-heating by the UVB is
balanced by cooling due to adiabatic expansion
(\citealt{HuiGnedin97,Valageas02}).  Measurements of the thermal state
of the IGM at $1.7<z<3.2$ from \Lya forest data give temperatures in
the range $10^{3.9}\rm~K<T_0<10^{4.4}\rm~K$ and are consistent with
slopes in the range of those expected within the standard paradigm for
the photoheated IGM, $1<\gamma<1.6$ (\citealt{Ricotti00,Schaye00,McDonald01}).    The fiducial model
(20-256) was chosen to be representative of these constraints.  In
order to explore the effect of different temperature-density relations
on the flux PDF, we also run six further simulations with different
thermal histories.  Four simulations (20-256t1, 20-256t2, 20-256g1 and
20-256g2) reproduce values of $T_{0}$ and $\gamma$ consistent with the
upper and lower end of the range quoted above.   We have also
simulated two further models  with $\gamma<1$ (20-256g3) and an
additional scatter in the temperature-density relation (20-256xs).  We
shall discuss the motivation behind these latter choices further in
section 3.   The different thermal histories are constructed by
modifying the fiducial simulation \HeII photo-heating rate,
$\epsilon_{\rm HeII}^{\rm fid}$, such that $\epsilon_{\rm HeII} =
\alpha \Delta^{\beta} \epsilon_{\rm HeII}^{\rm fid}$.  In the case of
the model with extra scatter (20-256xs), a random Gaussian distributed
dispersion is  applied  to the  \HeII photo-heating rate.  In this way
we generate  self-consistent thermal histories, in the sense that the
thermal state of the IGM is correctly coupled to the gas pressure.
Note, however, that a fully physically consistent treatment of the IGM
thermal history at $z\simeq 3$ would require cosmological radiative
transfer simulations with a level of detail beyond at least our
current numerical capabilities (although see
\citealt{Bolton04,Maselli05,Tittley07}). The values of  $\alpha$ and
$\beta$ used in all the simulations are summarised in Table 1, along
with the resulting parameters for the temperature-density relation in
each simulation. \footnote{The values of $\gamma$  in the fiducial
model are smaller than the value of $\gamma \simeq 1.6$ typically
reached under the assumption that the low density gas is optically
thin and in photoionisation equilibrium following reionisation.  We
have not rescaled the \HeII photo-heating rate in this instance.  This
difference is due to non-equilibrium ionisation effects which occur
during rapid changes in the UV background
(e.g. \citealt{MiraldaRees94,Haehnelt98,Theuns98}).  The smaller
values of $\gamma$ are due to the hardening of the UV background
spectrum around $z\simeq 3.5$ due to the increased contribution of
QSOs to the metagalactic UV background, leading to the reionisation
of the \HeII in the simulation.}

The cosmological parameters used for our fiducial simulation (20-256)
are consistent with the combined analysis of the third year WMAP and
\Lya forest data. $(\Omega_{\rm m},\Omega_{\Lambda},\Omega_{\rm
b}h^{2},h,\sigma_{8},n) = (0.26,0.74,0.024,0.72,0.85,0.95)$
(\citealt{Viel06,Seljak06}).  We also run three additional simulations
with different values for $\sigma_{8}$: one which is consistent with
the third year WMAP data alone (20-256s1, \citealt{Spergel07}) and
another two consistent with the somewhat higher constraints from weak
lensing and \Lya forest data (20-256s2 and  20-256s2g2,
\citealt{Lesgourgues07}).  The third model (20-256s2g2) also has a
slope for the temperature-density relation which is consistent with
the lower end of the observational constraints.  Two further models
were run using the multi-phase star formation model of
\cite{SpringelHernquist03}.  One model includes the multi-phase star
formation with winds disabled (20-256m), while the other includes the
effect of galactic winds with a velocity of $484 \rm~km~s^{-1}$
(20-256w).  This latter model is extreme in the sense that the kinetic
energy of the winds is comparable to the energy output from supernovae
in the model (\citealt{SpringelHernquist03b}) and should therefore
provide an upper limit on the impact galactic winds may have on the
PDF.

Lastly, to check numerical convergence we run five additional
simulations with different box sizes (40-200, 80-400) and mass
resolutions (20-100, 20-200, 20-400).  The parameters of these 
simulations are  listed in Table 1 along with a summary of all the
other models, while the  parameters characterising their resolution
and box-size are listed in Table 2.

\begin{table} 
\centering
  \caption{Mass resolution and box size of the five additional simulations
  used to check numerical convergence. The mass resolution and box
  size of the fiducial simulation is also listed for comparison at the
  bottom of the table.}
  \begin{tabular}{c|c|c|c}
    \hline
    Model     & Box size       & Total particle         & Gas particle  \\  
             & [comoving Mpc/h]   &  number  &  mass $[M_{\odot}/h]$ \\
  \hline
  20-100     & 20     & $2 \times 100^{3}$   & $1.03 \times 10^{8}$ \\
  20-200     & 20     & $2 \times 200^{3}$   & $1.29 \times 10^{7}$ \\
  20-400     & 20     & $2 \times 400^{3}$   & $1.61 \times 10^{6}$ \\
  40-200     & 40     & $2 \times 200^{3}$   & $1.03 \times 10^{8}$ \\
  80-400     & 80     & $2 \times 400^{3}$   & $1.03 \times 10^{8}$ \\
   \hline
  20-256     & 20     & $2 \times 256^{3}$   & $6.13 \times 10^{6}$ \\
  \hline
\end{tabular}
\end{table}

\subsection{Synthetic \Lya spectra}

Sets of synthetic \Lya spectra are constructed at $z=2.07$, $z=2.52$
and $z=2.94$ using 1024 random lines of sight drawn from the
simulations (e.g. \citealt{Theuns98}).  The line profile convolution
is performed with  the Voigt profile approximation of
\cite{TepperGarcia06}.  We rescale the synthetic spectra to match the
mean normalised flux, $\langle F \rangle$, in the redshift bins used
by K07 to measure the flux PDF: $\langle F \rangle =
[0.863,0.797,0.730]$ at $\langle z \rangle =[2.07,2.52,2.94]$.  These
values correspond to an effective optical depth, $\tau_{\rm eff} =
-\ln(\langle F \rangle) = [0.147,0.227,0.315]$.

The raw spectra are then processed to resemble the observational data.
The spectra are convolved with a Gaussian with $\rm FWHM = 7~\rm
km~s^{-1}$ and are rebinned onto pixels of width $0.05\rm~\AA$.  Noise
is then added to the spectra in a similar manner to \cite{Rauch97} and
\cite{McDonald00}.  Within each of the three observationally defined
redshift bins for the PDF, the variance of the noise in the observed
spectra is determined over flux ranges corresponding to each flux bin
of the PDF.  Note that we have used the same binning for the flux PDF
as \cite{McDonald00} and K07.  The PDF bins have a width of $\Delta
F=0.05$, with the first and last bins centred on $F=0$ and $F=1$
respectively.  This gives $21$ bins in total from $F=0$ to $F=1$.
Gaussian distributed noise consistent with the variances in each flux
bin of the observational data is then added to the corresponding
pixels in the synthetic spectra.  This process mimics the noise
properties of the observed spectra well (\citealt{Rauch97}). Note that
all pixels with flux levels smaller than $F=0.025$ or greater than
$F=0.975$ have been allocated to the $F=0$ or $F=1$ PDF bins,
respectively.

\section{Comparison of the K07 PDF to hydrodynamical simulations}
\subsection{The effect of resolution, box-size and galactic winds}

\begin{figure*}
\centering 
\begin{minipage}{180mm} 
\begin{center}

\psfig{figure=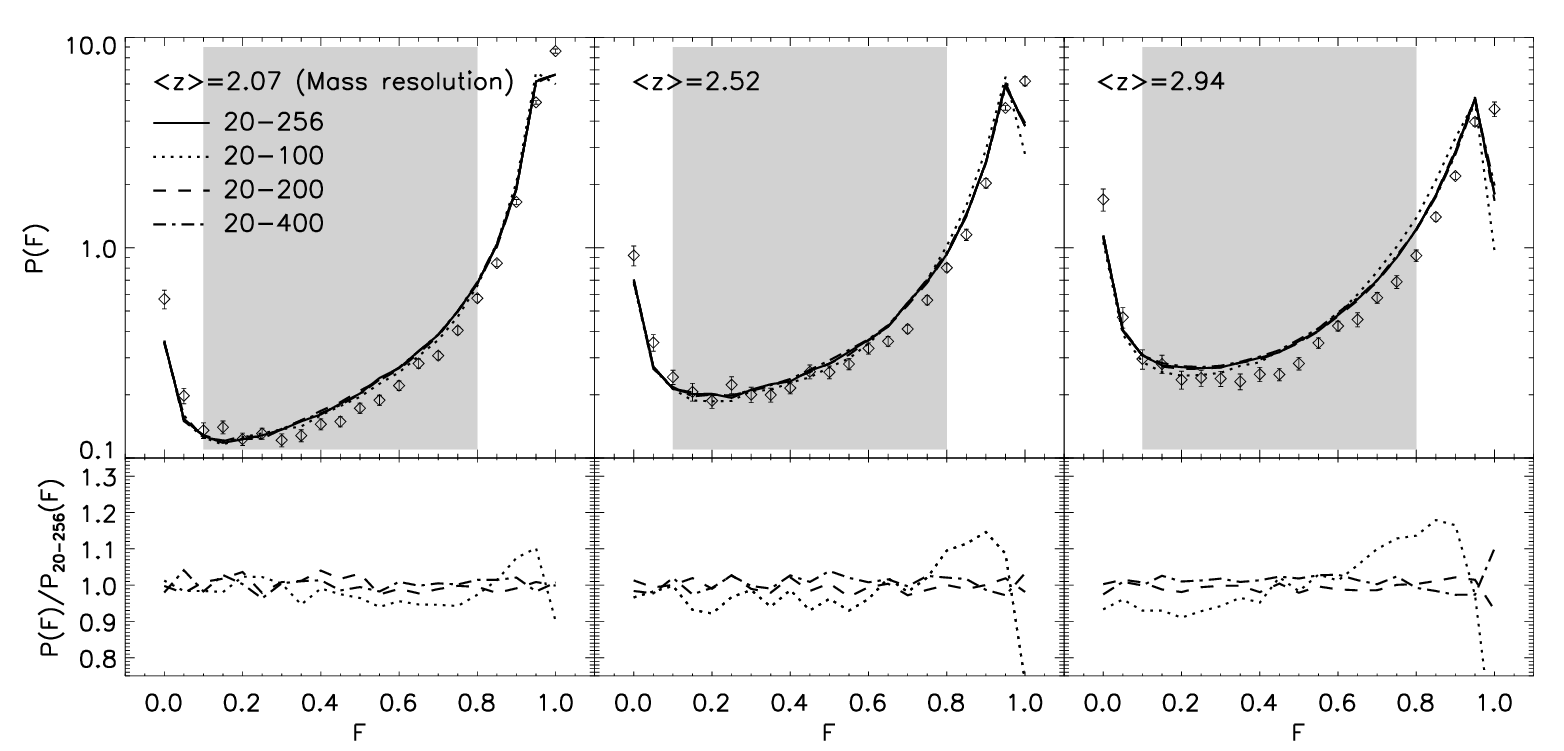,width=0.8\textwidth}
\vspace{-0.2cm}
\psfig{figure=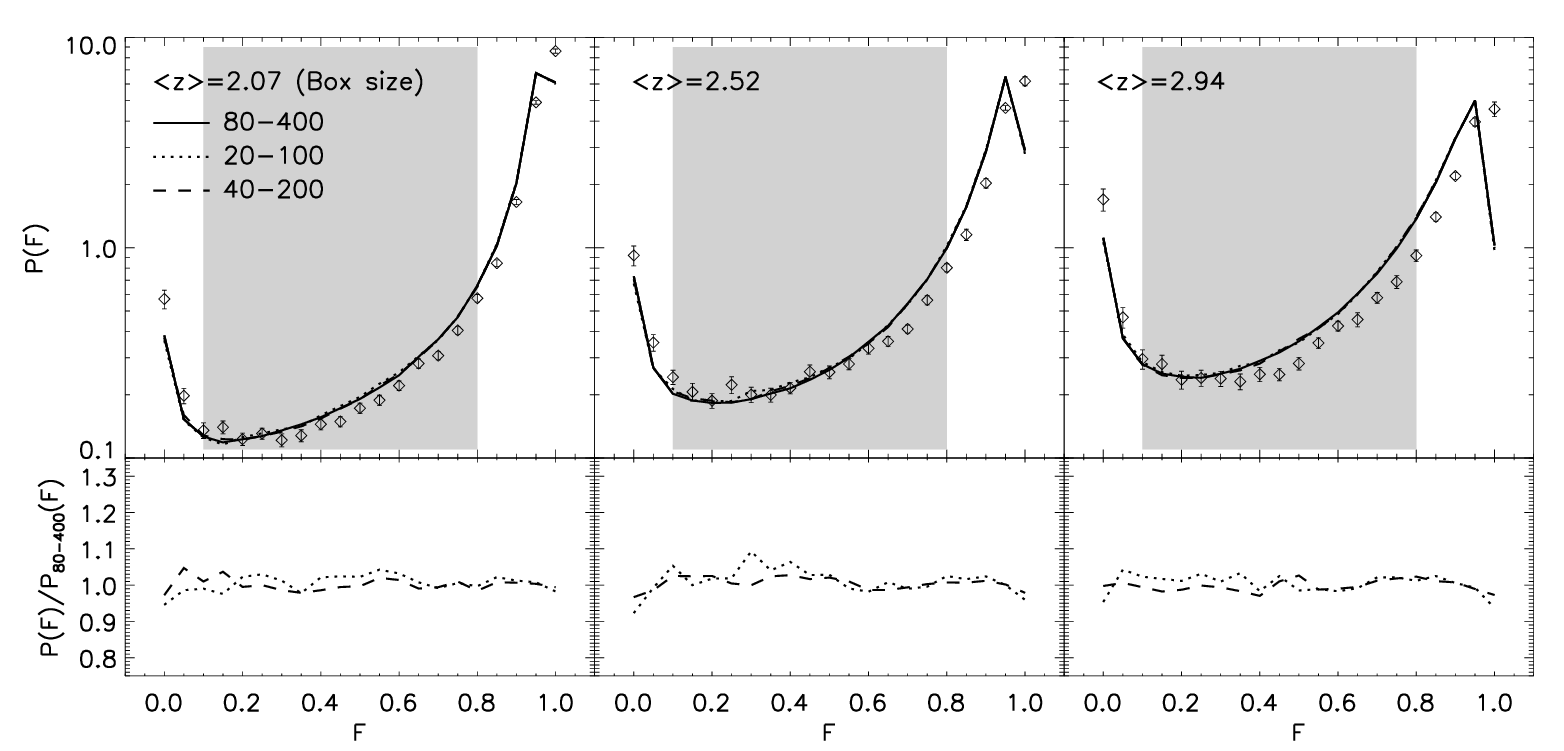,width=0.8\textwidth}
\vspace{-0.2cm}
\psfig{figure=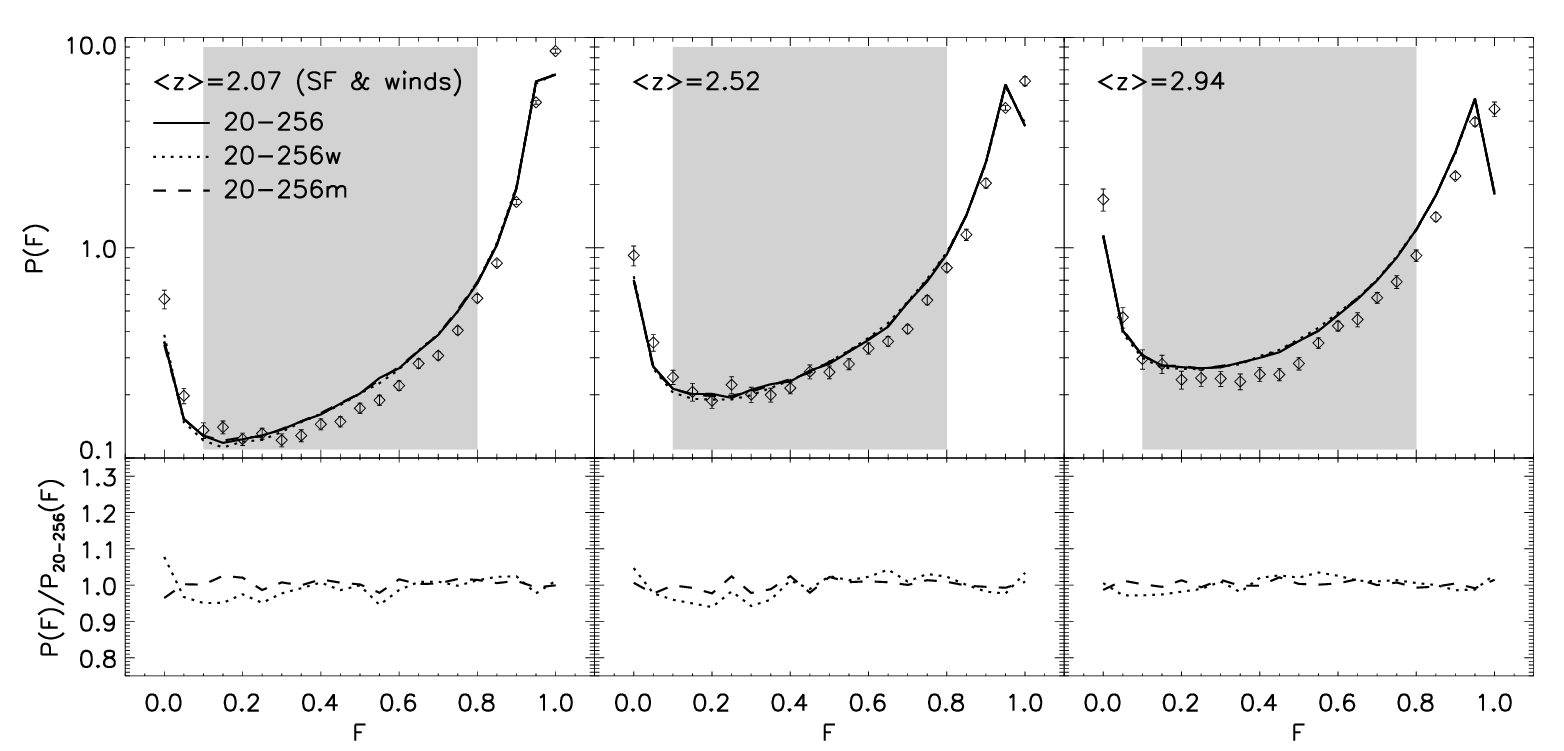,width=0.8\textwidth}

\caption{The impact of resolution, box-size and galactic winds on the simulated
flux PDF.  {\it Top row:} The effect of mass resolution on the
simulated flux PDF. The lower section in each panel displays the ratio
of the flux PDFs to the fiducial model (20-256). {\it Middle row:}
The effect of box size. The lower section in each panel displays the
ratio of the flux PDFs to the model with the largest box size
(80-400). {\it Bottom row:}  The effect of the adopted star formation
model and galactic winds. The lower section in each panel again
displays the ratio of the flux PDFs to the fiducial model (20-256).
For comparison, the observational data of K07 are also shown  (open
diamonds with error bars) and the shaded grey region in each panel
corresponds to the portion of the PDF which is least affected by noise
and continuum uncertainties.}

\label{fig:numerical} 
\end{center} 
\end{minipage}
\end{figure*}

Fig.~\ref{fig:numerical} displays the effect of resolution, box-size 
and galactic winds on the simulated flux PDF. The observational data of K07
are shown by the open diamonds with error bars, and the shaded regions
in all panels correspond to the flux range over which noise properties
and continuum placement have a small impact on the PDF, $0.1 \leq
F \leq 0.8$ (K07). 

The top row in Fig.~\ref{fig:numerical} displays the effect of mass
resolution on the simulated PDF.  The ratio of the three simulated
PDFs with different mass resolutions to the fiducial model PDF
(20-256, see Table 2 for details) is shown in the lower section in
each of the panels.  The simulated PDF appears to be marginally
converged and is generally to within a few percent of the highest
resolution run at all redshifts.  As we shall see in the next section,
these differences are small compared to those exhibited when other
parameters are varied, most notably $\sigma_8$ and $\gamma$.  Note,
however, the flux PDF of the fiducial model does not agree well with
the PDF from the observational data of K07.  We shall return to this
point later in this paper.

In the middle row, the effect of the simulation box size on the PDF is
displayed.  The lower sections of each panel show the ratio of the
PDFs from the two smaller boxes to the PDF extracted from the largest
box (80-400).  All three simulations have the same mass resolution.
Again, the differences in the simulated PDFs are very small.  We
therefore consider our fiducial simulation box size and mass
resolution to be adequate for distinguishing between the various
IGM models considered in this work.

The bottom row of panels in Fig.~\ref{fig:numerical} shows a
comparison of the PDFs computed from the fiducial model (20-256, solid
curve) to the results from the simulations including the multiphase
star formation (SF) model of \cite{SpringelHernquist03} with (20-256w,
dotted curve) and without (20-256m, dashed curve) winds.  The lower
sections in each panel again display the ratio of these PDFs to the
fiducial model.  The differences between the fiducial and multiphase
SF model PDFs are within a few percent, a result which gives us
confidence the simple star formation prescription used for the
majority of the simulations in this work has little impact on the
properties of the \Lya forest.  This is because the highly overdense
regions containing most of the star formation correspond to only a
very small proportion of the volume probed by the synthetic spectra.
The difference between the model with strong galactic winds is also
small, except at $F=0$ in the $\langle z \rangle = 2.07$ bin.
However, this is a rather extreme model and may overestimate the
impact of winds on the \Lya forest.  In comparison, after rescaling
the mean flux of their synthetic spectra, \cite{Theuns02b} find winds
also have a very small effect on the \Lya forest.  A similar result
was found by \cite{Bruscoli03}.  \cite{BertoneWhite06} used a
semi-analytical model to investigate the impact of galactic winds on
the \Lya forest, and also found the impact of winds on the PDF is
negligible.  Our results are in qualitative agreement with these
studies, and we  expect the effect of galactic winds on the \Lya flux PDF
to be small.  Although we have not analysed the possible impact of
outflows from rarer, more massive galaxies hosting active galactic
nuclei (AGN) on the \Lya forest, these are expected to extend into a
smaller volume of the IGM than the supernova driven galactic winds
(\citealt{Sijacki07}).  AGN feedback is therefore also likely to have
a negligible impact on the \Lya forest flux distribution.

\subsection{The effect of cosmological and astrophysical parameters} \label{sec:params}

\begin{figure*}
  \centering 
  \begin{minipage}{180mm} 
    \begin{center}
            	
      \psfig{figure=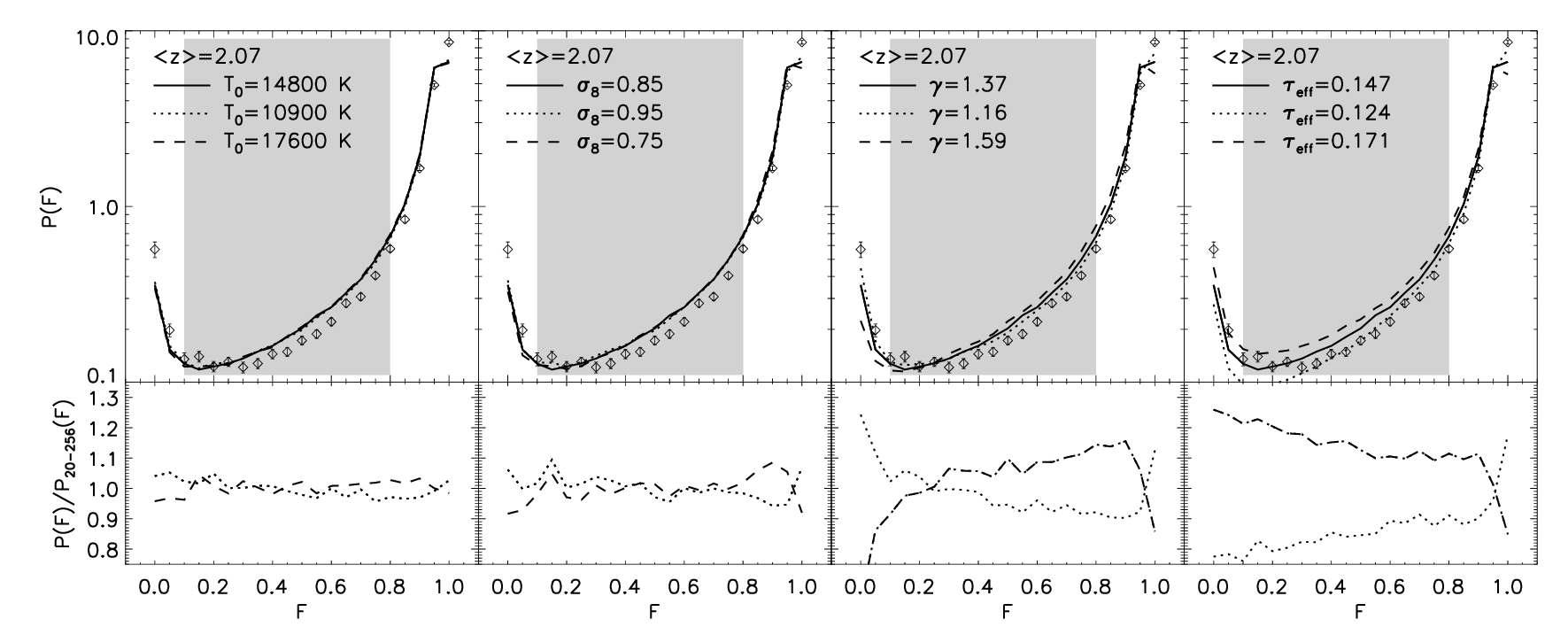,width=0.95\textwidth}
      \psfig{figure=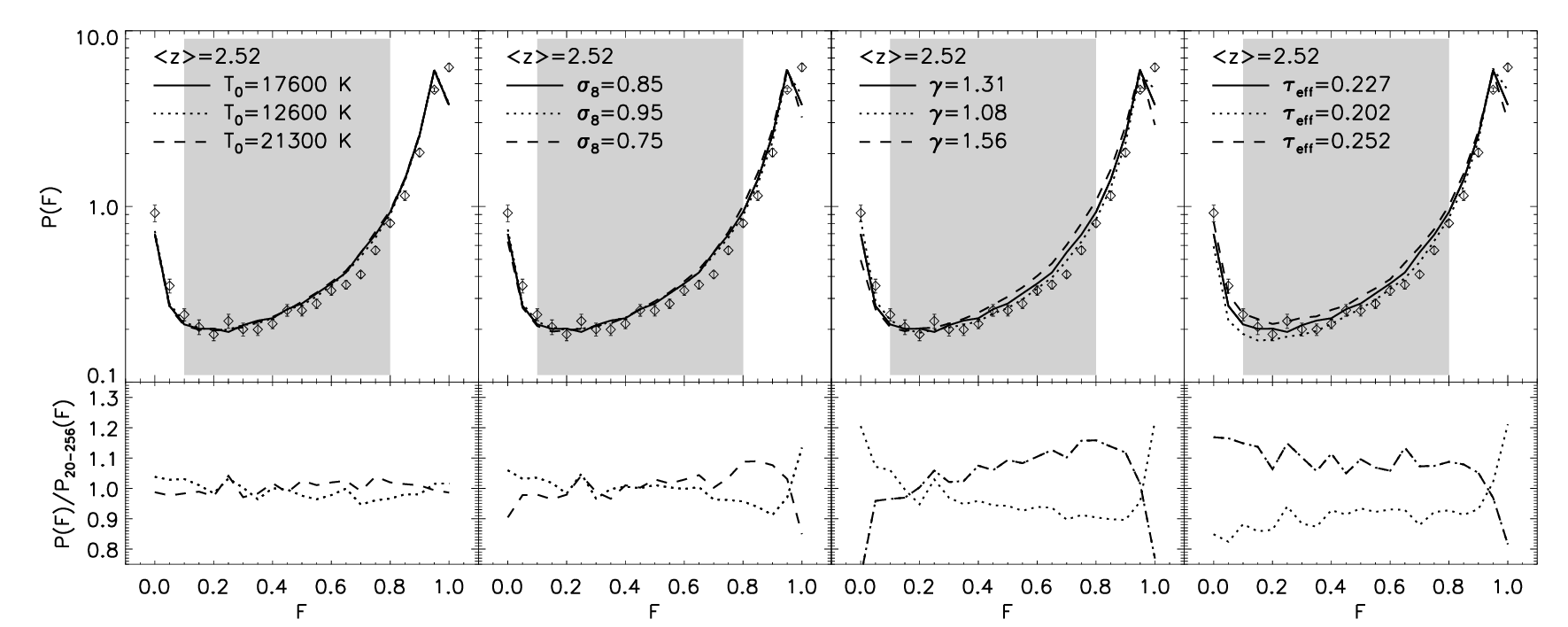,width=0.95\textwidth}
      \psfig{figure=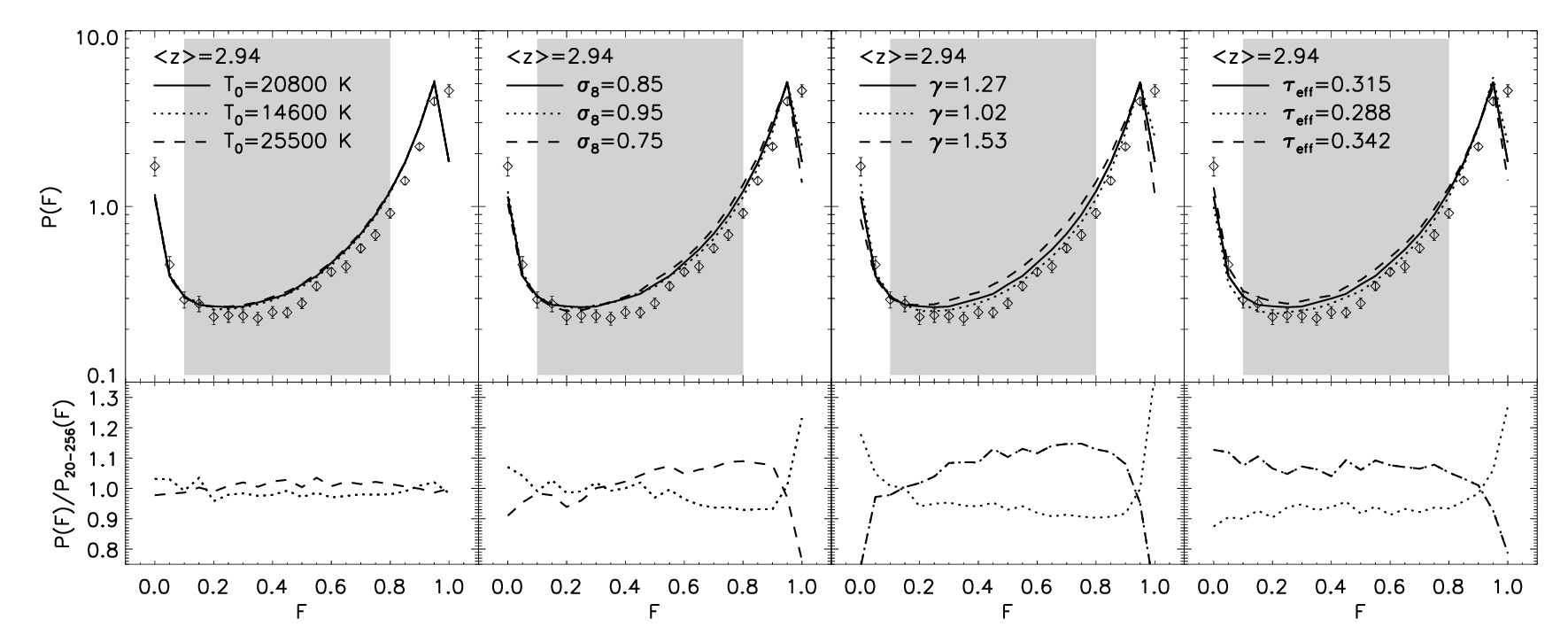,width=0.95\textwidth}

      \caption{Comparison of the flux PDF computed from hydrodynamical
      \Lya forest simulations with different cosmological and
      astrophysical parameters to the observational data of K07. {\it
      Top row:} The observed PDF at $\langle z \rangle =2.07$ (open
      diamonds with error bars).  The solid curve in each panel shows
      the PDF derived from the fiducial model. From left to right, the
      dotted and dashed curves demonstrate the effect on the PDF of
      varying $T_0$, $\sigma_8$, $\gamma$ and $\tau_{\rm eff}$ within
      the observationally determined boundaries.  The lower section of
      each plot shows the ratio of these PDFs to the fiducial model.
      {\it Middle Row:} As for top row but at  $\langle z \rangle
      =2.52$. {\it Bottom Row:} As for top row but at $\langle z
      \rangle =2.94$.  The shaded grey region in each panel
      corresponds to the portion of the PDF which is least affected by
      noise and continuum uncertainties.}  \label{fig:pdf_all}
      \end{center} \end{minipage}
\end{figure*}

The flux PDFs constructed from the simulations with different
cosmological and astrophysical parameters are compared to the
observational data of K07 (open diamonds with error bars) in
Fig.~\ref{fig:pdf_all}.  From left to right, in each of the four
panels the effect of varying $T_0$, $\sigma_8$, $\gamma$ and
$\tau_{\rm eff}$ on the PDF is shown.  The solid curve in each panel
corresponds to the PDF of the fiducial model (20-256).  The lower
segment of each panel shows the ratio of the PDFs indicated in the
diagram to the fiducial model. Note again that except where indicated
in the rightmost panels of Fig.~\ref{fig:pdf_all}, all the synthetic
spectra are rescaled to have  $\tau_{\rm eff} = [0.147,0.227,0.315]$
at $\langle z \rangle =[2.07,2.52,2.94]$, corresponding to the
effective optical depth in the redshift bins for which K07 have
reported their measurement of the flux PDF.

At all redshifts the PDF is relatively insensitive to changing the IGM
temperature at mean density, $T_0$, for  values within the
observationally constrained range (\citealt{Schaye00}).  Higher
temperatures slightly increase the amplitude of the PDF at $0.1<F<0.9$
and decrease the amplitude at $F=0$ and $F=1$.  This behaviour is
attributable to the Jeans smoothing effect of gas pressure on the
spectra; higher temperatures increase the IGM gas pressure and smooth
the IGM over larger scales.  Combined with the increased Doppler
widths in the higher temperature models, this redistributes the PDF by
effectively erasing the small scale power in the spectra.  Varying 
$\sigma_8$ has a slightly larger effect.  Models with smaller values
of $\sigma_8$ tend to
produce broader absorption lines which arise from the slightly
smoother gas distribution at scales comparable to the Jeans length.
The effect is to again erase small scale power.  
The effect of $\sigma_8$ and temperature  on the flux PDF 
are, however,  small and not sufficient to account for the differences 
between the K07 data and the simulations. A more detailed
discussion of the effect of these parameters on flux statistics can be
found in \cite{Theuns00}.

Varying the temperature-density relation index, $\gamma$, has a much 
stronger effect on the PDF. Altering $\gamma$ changes the temperature
of the IGM in a density dependent fashion; decreasing $\gamma$
increases the IGM temperature below the mean IGM density and decreases
the temperature above mean density.  If we assume  \HI to be in
photoionisation equilibrium, the neutral hydrogen density scales
approximately as $n_{\rm HI} \propto T_{0}^{-0.7} \Gamma_{\rm HI}^{-1}
\Delta^{2-0.7(\gamma-1)}$, where $\Gamma_{\rm HI}$ is the
photoionisation rate per \HI atom (e.g.  \citealt{Rauch98}).  
Note that this is not the case for changes in $T_0$ only, where the
gas temperature (and hence also the residual \HI fraction in the IGM)
are altered independently of the gas density.

These combined effects explain the changes seen in the PDF for
different $\gamma$.  As shown in Fig.~\ref{fig:pdf_all}, a lower value
of $\gamma$ decreases the amplitude of the PDF moderately over the
flux range $0.1<F<0.9$, an effect which is compensated for by a
strong increase of the amplitude over the flux ranges at $F=0$
and $F=1$.  Absorption lines associated with moderately overdense
regions become associated with colder, more neutral material and hence
are narrower and have a higher opacity in the centre of the line. In
contrast, underdense regions associated with transmission near the
continuum level become hotter and therefore more highly ionised.
Expressed in a different way, a larger $\gamma$ tends to erase the
small scale power in the spectra by increasing the temperature of the
gas associated with the most prominent absorption lines.  Note that the
effect of changing $\gamma$ is therefore partially degenerate with
changing $T_0$ and $\sigma_8$ within the observationally constrained
ranges, although it has a significantly larger impact on the shape of
the PDF.   Fig.~\ref{fig:pdf_all}  suggests  that a
temperature-density relation which is close to isothermal is in better
agreement with the K07 data.  

Finally, the effect of varying $\tau_{\rm eff}$ on the PDF is
illustrated in the right-most panels of Fig.~\ref{fig:pdf_all}.  The
value of  $\tau_{\rm eff}$  measured in observed spectra exhibits
rather large fluctuations between different lines of sight, and it is
rather sensitive to the number of strong absorption systems
(\citealt{Viel04}) and the contribution of metal absorption (K07).  It
is therefore reasonable to consider $\tau_{\rm eff}$ as an independent
free parameter within the measurement errors of  the observations.
The values displayed correspond to a mean flux range of $\pm 0.020$
around the fiducial values of $\langle F \rangle =
[0.863,0.797,0.730]$ at $\langle z \rangle =[2.07,2.52,2.94]$.  The
PDF is very sensitive to  even small changes of $\tau_{\rm eff}$  but
in a qualitatively different way to the other parameters. Altering
$\tau_{\rm eff}$  changes the shape of the PDF by increasing the PDF
amplitude at one end and decreasing it at the other. This is in
contrast to the effect of the other parameters, which when varied
raise or lower the PDF at both ends simultaneously.

\subsection{Radiative transfer effects during reionisation}

\begin{figure*}
\centering 
\begin{minipage}{180mm} 
\begin{center}
\psfig{figure=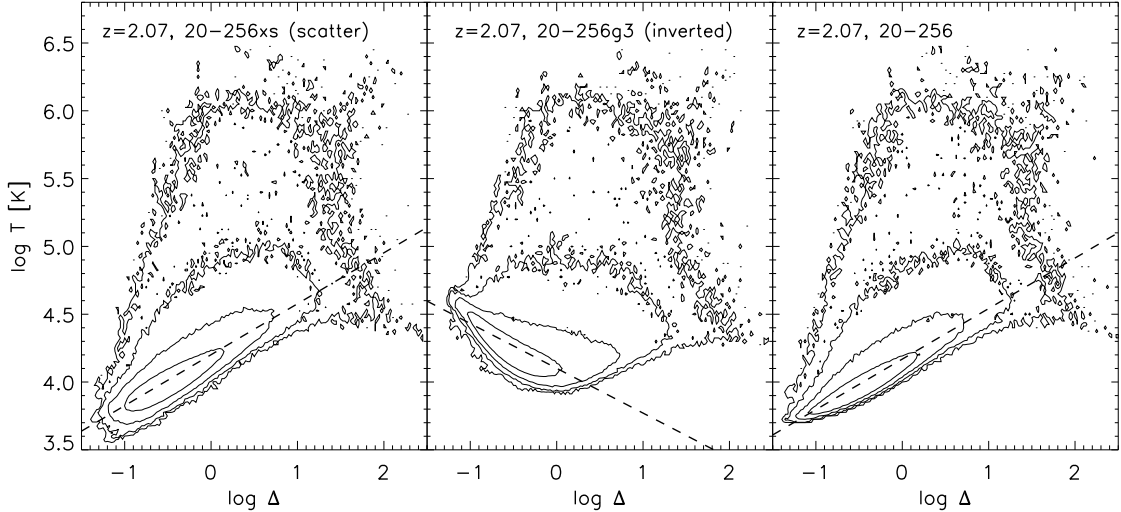,width=0.85\textwidth}
\caption{Contour plots of the volume weighted temperature-density
plane at $z=2.07$ in three of our hydrodynamical simulations. The
number density of the data points increases by an order of magnitude
with each contour level.  The model in the left panel (20-256xs) has a
temperature-density relation which exhibits a moderately increased
scatter, while the central panel displays the simulation (20-256g3)
with an inverted temperature-density relation.  For comparison, the
temperature-density plane in the fiducial model (20-256) is shown in
the right hand panel.  The dashed lines in each panel correspond to
the temperature-density relations listed in Table~$1$.}
\label{fig:temprho} 
\end{center} 
\end{minipage}
\end{figure*}

Thus far we have assumed a tight, polytropic temperature-density
relation for the low density IGM, $T = T_{0}\Delta^{\gamma-1}$, with
$1\le \gamma\le1.6$, which in the standard paradigm for the  thermal
state of the photo-ionised IGM  is expected to apply to the majority
of the optically thin gas responsible for the absorption in the \Lya
forest.  The relation arises from the balance between photo-heating
and cooling by adiabatic expansion, with a slope which evolves from
$\gamma \simeq 1$ following reionisation to an asymptotic value of
$\gamma \simeq 1.6$ (\citealt{HuiGnedin97,Theuns98,Valageas02}).
Observational evidence for a boost in the IGM temperature and/or a
flattening in the temperature-density relation at $z\simeq 3$ provides
possible evidence for \HeII reionisation
(\citealt{Schaye00,Ricotti00,Theuns02,Zaroubi06}).  As we saw in the
last section, the observed flux PDF indeed appears to prefer a thermal
state where the voids in the IGM are relatively hot, although this
model is still not in perfect agreement with the data.  It is
therefore possible that a model with an inverted temperature-density
relation with $\gamma<1$ may instead provide a better fit to the data.

Is such a thermal state for the IGM plausible? During reionisation the
assumption of a tight temperature-density relation for the low density
IGM is likely to be a poor one.  The filtering of ionising radiation
by optically thick gas alters the spectral shape of the UV background,
giving rise to additional heating effects which raise the gas
temperature and blur the dependence of the IGM temperature on the gas
density (\citealt{AbelHaehnelt99,Bolton04,Maselli05,Tittley07}).  The
inhomogeneous distribution of the ionising sources themselves will
also contribute to spatial variations in the IGM thermal  state
(\citealt{MiraldaEscude00,Gleser05,Furlanetto07}).  In addition, the
cooling timescale for the low density gas responsible for the \Lya
forest is of order a Hubble time, meaning that changes to the thermal
state of the IGM will persist for some time after reionisation
(\citealt{MiraldaRees94}).

Radiative transfer effects on the thermal state of the IGM may thus
result in  an inversion of the temperature-density relation, such that
$\gamma<1$.  As ionising photons propagate through optically thick,
moderately overdense regions in the IGM, the spectral shape of the
emission is considerably hardened as lower energy photons are
preferentially absorbed.  Any subsequent underdense gas is then
subjected to ionising photons which have a substantially higher mean
energy, boosting the temperature of the underdense region as it is
photoionised and heated.  One may envisage such a situation if
ionising sources are initially embedded in overdense, optically thick
regions.  State-of-the-art cosmological radiative transfer simulations
indicate that this ``inside-out'' topology may be appropriate for
moderate over and underdensities, at least during \HI reionisation
(\citealt{Iliev06b}).  Such a radiative transfer induced temperature
inversion effect was noted by \cite{Bolton04}, although this result
was a specific case for a single density distribution.  More
generally, radiative transfer effects during inhomogeneous
reionisation are expected to produce a more complex, multiple valued
relationship between temperature and density in the IGM.  While we
were in the final stages of preparing this manuscript
\cite{Furlanetto07} presented a semi-analytical model aimed at
describing  the thermal state of the IGM during  and after the
inhomogeneous reionisation of helium. While probably not fully
realistic due to a lack of a proper modeling of the filtering and
spectral hardening effects of radiative transfer, the model confirms
that an initially inverted and multi-valued temperature
density-relation is indeed plausible.  The exact character of these
effects will also depend on the types of sources which reionise the
IGM (\citealt{Tittley07}).

\begin{figure*}
\centering 
\begin{minipage}{180mm} 
\begin{center}
\psfig{figure=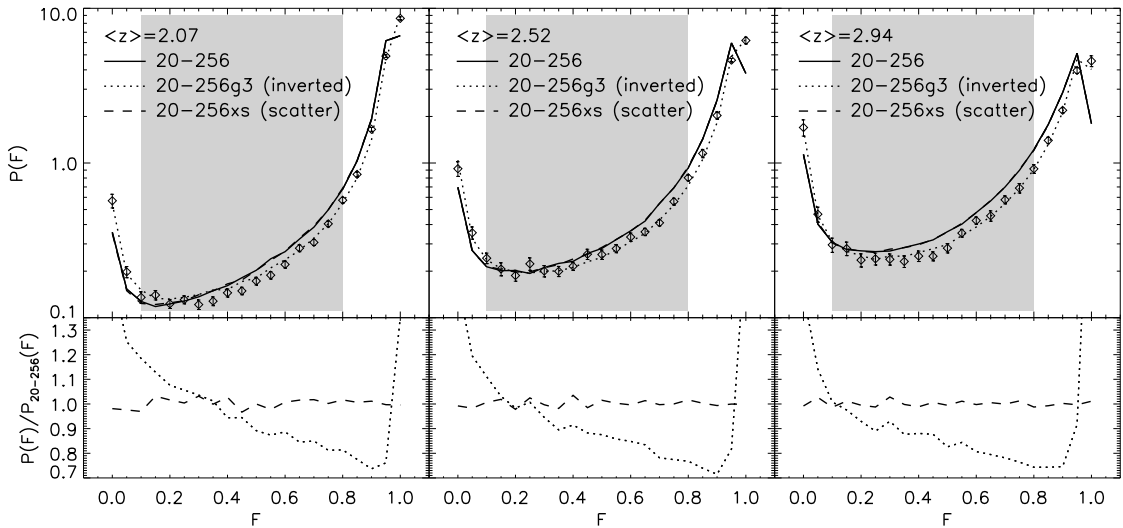,width=0.85\textwidth}
\caption{Comparison of the observational data of K07 at $\langle z
\rangle =[2.07,2.52,2.94]$ (open diamonds with error bars) to 
simulations with an inverted temperature-density relation (20-256g3,
dotted curve) and a model with increased scatter (20-256xs, dashed
curve). The solid curve in each upper panel shows the PDF derived from
the fiducial model (20-256), while the lower panels display the ratio
of the other PDFs to the fiducial model.  The shaded grey region in
each panel corresponds to the portion of the PDF which is least
affected by noise and continuum uncertainties.}
\label{fig:gammainv} 
\end{center} 
\end{minipage}
\end{figure*}

Consequently, adopting a temperature-density relation with $\gamma<1$
may mimic some of the different heating effects expected due to
inhomogeneous radiative transfer through an optically thick IGM. One
may expect many underdense regions to exhibit significantly elevated
temperatures following reionisation
(\citealt{AbelHaehnelt99,Bolton04,Gleser05,Tittley07}). Helium in the
IGM at $z \simeq 3$ exhibits patches which are still optically thick
to photons with energies in excess of $4\rm~Ry$
(e.g. \citealt{Jakobsen94,Heap00,Reimers05b}).  Radiative transfer
effects on the \Lya flux PDF may therefore be important at the
redshifts we consider here, especially if the tail end of \HeII
reionisation occurs at $z\simeq 3.2$ (e.g.
\citealt{Songaila98,Schaye00,Theuns02,Bernardi03,Shull04,Bolton06,Faucher07}).
Furthermore, using a semi-analytical model for the \Lya forest based
on density distributions drawn from hydrodynamical simulations,
\cite{Becker07} noted that an inverted temperature-density relation
improved their fit to the flux PDF measured with an independent set of
high resolution QSO spectra at $1.7<z<5.8$.

The inhomogeneous spatial distribution of ionising sources is, however,
also expected to increase the {\it scatter} in the temperature-density
relation (\citealt{Theuns02c}).  We therefore compare two further
different models to the K07 data, one with an inverted
temperature-density relation (20-256g3) and another with a moderately
increased scatter in the temperature-density relation (20-256xs).  For
illustration,  the volume weighted temperature-density planes for
these two  models at $z = 2.07$ are displayed in
Fig.~\ref{fig:temprho}. The fiducial model is shown in the right-most
panel.  The dashed lines in each panel correspond to the
temperature-density relations listed in Table~$1$.

Fig.~\ref{fig:gammainv} displays the comparison between the flux PDFs
constructed from these models to the observational data. Note  that we
have again matched $\tau_{\rm eff}$ in the simulated spectra  to the
values measured by  K07.   The PDFs for the fiducial model (solid
curves) and the model with increased scatter in the
temperature-density relation  (dashed curves) are very similar. At all
redshifts the flux PDFs exhibit differences of only a few per cent.  A
moderate increase in the scatter of the temperature-density relation
appears unlikely to substantially affect the shape of the PDF.
However, the temperature-density relation is still relatively well
defined in this model.  There is also no good reason to assume a
Gaussian dispersion. In reality the scatter may be rather more
inhomogeneous.  We therefore should not rule out the possibility that
more complex, multiple valued relationship between temperature and
density fits the K07 data better.  We now turn to the model with an
inverted temperature-density relation (dotted curves). In this case
the modification of the thermal state of the IGM makes a substantial
difference to the shape of the PDF, as expected from  the behaviour
seen in Fig.~\ref{fig:pdf_all}.  The inverted temperature-density
relation brings the simulated and observed flux PDFs into
excellent agreement.  The previously most  problematic points  at the
extreme ends of the PDF at flux levels close to  zero and  unity now
show an excellent match between the simulations and data.  Note again,
however, that the IGM may nevertheless have a more complex thermal
state following \HeII reionisation, rather than actually having a
well defined, inverted temperature-density relation.  We now proceed
to discuss the quality of the fits of different models more
quantitatively  in the next section.

\section{Determination of the best fitting models to the K07 PDF}
\subsection{A quantitative comparison of the simulated and observed data}
  
As should be clear from the previous section and analyses
by other authors, there is considerable degeneracy  in the dependence
of the flux PDF on cosmological and astrophysical parameters. The
quality of the observational data and  the apparently good agreement
with the flux PDF from our inverted temperature-density relation
simulation motivated  us, however, to perform a quantitative
$\chi^{2}$ analysis. For this we used the full covariance matrix
derived from the observational data of K07.

\begin{table*} 
\centering 
\caption{The minimum $\chi^{2}$ values  for the best
  fitting models to the K07 PDF data at $\langle z \rangle =
  [2.07,2.52,2.94]$, fitted
  in the range $0.1 \leq F \leq 0.8$ only.  The columns list the best
  reduced $\chi^{2}$ (for $\nu=15-4=11$ degrees of freedom) and the
  corresponding best fitting value for the effective optical depth,
  $\tau_{\rm fit}$.  This is expressed as a ratio with the mean value
  of $\tau_{\rm eff}$ in each observational redshift bin.  For
  reference these are $\tau_{\rm eff} = [0.147,0.227,0.315]$ at
  $\langle z \rangle = [2.07,2.52,2.94]$.}

  \begin{tabular}{c|c|c|c|c|c|c|c}
    \hline
    Model   & $\chi^{2}/ \nu$ & $\chi^{2}/ \nu$ & $\chi^{2}/ \nu$ &
    $\tau_{\rm fit}/ \tau_{\rm eff}$ &  $\tau_{\rm fit}/ \tau_{\rm
    eff}$ &  $\tau_{\rm fit}/ \tau_{\rm eff}$ & Notes \\  
   & ($z=2.07$) & ($z=2.52$) & ($z=2.94$) & ($z=2.07$) &
    ($z=2.52$) &  ($z=2.94$) &  \\
  \hline
  20-256     & 4.57 & 2.84 & 2.03 & 0.840 & 0.857 & 0.817 & Fiducial
    model \\
  20-256t1   & 3.61 & 2.49 & 1.86 & 0.864 & 0.871 & 0.822 & Low $T_{0}$\\
  20-256s2   & 4.05 & 2.65 & 1.65 & 0.886 & 0.851 & 0.819 & $\sigma_8=0.95$ \\
  20-256g2   & 2.34 & 1.90 & 1.34 & 0.898 & 0.927 & 0.849 & Low $\gamma$\\
  20-256s2g2 & 1.85 & 1.81 & 1.14 & 0.888 & 0.925 & 0.923 & $\sigma_8=0.95$, Low $\gamma$ \\
  20-256g3   & 1.62 & 1.44 & 0.90 & 0.935 & 1.025 & 1.013 & Inverted $\gamma$ \\

  \hline
\end{tabular}
\end{table*}

\begin{table*} 
\centering

  \caption{The $\chi^{2}$ analysis results for the best fitting models
  to the whole PDF, $0 \leq F \leq 1$.  As in Table 3, the columns
  list the best reduced $\chi^{2}$ (for $\nu=21-4=17$ degrees of
  freedom) and the corresponding best fitting value for the effective
  optical depth, $\tau_{\rm fit}$, as a ratio of the mean observed
  value in each redshift bin.  For reference these are $\tau_{\rm eff}
  = [0.147,0.227,0.315]$ at $\langle z \rangle = [2.07,2.52,2.94]$.}

  \begin{tabular}{c|c|c|c|c|c|c|c}
    \hline
    Model   & $\chi^{2}/ \nu$ & $\chi^{2}/ \nu$ & $\chi^{2}/ \nu$ &
    $\tau_{\rm fit}/ \tau_{\rm eff}$ &  $\tau_{\rm fit}/ \tau_{\rm
    eff}$ &  $\tau_{\rm fit}/ \tau_{\rm eff}$ & Notes  \\  
   & ($z=2.07$) & ($z=2.52$) & ($z=2.94$) &  ($z=2.07$) &
    ($z=2.52$) &  ($z=2.94$) & \\
  \hline
  20-256     & 11.15 & 15.28 & 13.47 & 0.781 & 0.833 & 0.823 & Fiducial
    model\\
  20-256t1   & 10.02 & 15.22 & 14.14 & 0.808 & 0.840 & 0.844 & Low $T_{0}$\\
  20-256s2   & 6.90  & 9.65  & 9.62  & 0.820 & 0.850 & 0.854 & $\sigma_8=0.95$\\
  20-256g2   & 5.03  & 8.14  & 8.03  & 0.884 & 0.914 & 0.902 &  Low $\gamma$\\
  20-256s2g2 & 3.07  & 4.70  & 5.32  & 0.923 & 0.930 & 0.927 &
    $\sigma_8=0.95$, Low $\gamma$ \\
  20-256g3   & 3.06  & 1.56  & 1.82  & 1.057 & 1.029 & 1.025 & Inverted $\gamma$ \\

  \hline
\end{tabular}
\end{table*}

As discussed by \cite{McDonald00}, using the covariance matrix is
important; the errors bars on the PDF are highly correlated.  However,
the covariance matrices derived by K07 (their equation 1) are rather
noisy for this purpose, preventing reliable inversion.  We therefore
follow the method suggested by \cite{Lidz06b}, where the observed
covariance matrix is regularised using the correlation coefficients,
$r_{\rm s}(i,j) = cov_{\rm s}(i,j)/ \sqrt{cov_{\rm s}(i,i)cov_{\rm
s}(j,j)}$ computed for each set of synthetic spectra, such that
$cov_{\rm d}(i,j)=r_{\rm s}(i,j)\sqrt{cov_{\rm d}(i,i)cov_{\rm
d}(j,j)}$.  Here $cov_{\rm d}$ and $cov_{\rm s}$ are the covariance
matrices computed from the observed data and synthetic spectra,
respectively.  Note that this implicitly assumes the observed and
simulated spectra have similar covariance properties.

We perform a simple $\chi^{2}$ analysis on the K07 data for two
different cases.  In the first instance we adopt a conservative
approach and analyse the PDF only over the flux range $0.1 \leq F \leq
0.8$, where the  impact of noise properties and continuum
uncertainties on the K07 PDF data are small.  This is similar to the
approach adopted by \cite{Desjacques05}.  However, as demonstrated
in the last section, the extreme ends of the PDF at low and high flux
levels contain important information on the thermal state of the IGM.
For  high resolution, high $S/N$ data like that of K07, the
uncertainties are expected to be relatively small over the entire
range of the flux PDF.  The continuum fits are generally accurate to
within $1-2$ per cent, and the typical $S/N\simeq 50$ (see K07 for a
detailed discussion regarding the effect of these parameters on the
PDF).  We therefore also perform the $\chi^{2}$ analysis over the
entire PDF,  $0 \leq F \leq 1$.

In our analysis we do not attempt a global minimisation  of $\chi^{2}$
over the parameters we have varied in the simulations.  We instead
just tabulate the minimum reduced $\chi^{2}$ values for our grid of models with
different  $T_0$, $\sigma_8$, and $\gamma$.  We  vary $\tau_{\rm eff}$
over an appropriate range, 25 per cent either side of the central
values from the fit to the redshift evolution of the metal cleaned \Lya
effective optical depth reported by K07, $\tau_{\rm eff} = 0.0023 \pm
0.0007(1+z)^{3.65 \pm 0.21}$.  This scatter reflects the variance in
the data.  Continuum fitting uncertainties, although small at the
redshifts we consider here, are another source of error in constraints
on this quantity. We shall discuss this in more detail shortly.

The results of the $\chi^{2}$ analysis for the flux range $0.1\leq F
\leq 0.8$ are summarised in Table 3.  In addition to the fiducial
model (20-256) we tabulate the results for five other simulations with
varying $T_0$, $\gamma$ and $\sigma_8$.   The models which have a low
$\sigma_8$, high $T_0$, high $\gamma$ and increased scatter in the
temperature-density relation are omitted from Table$~3$, as all have
reduced $\chi^2$ values which are similar to or worse than the
fiducial model.  The inverted temperature-density relation model
(20-256g3) provides the best fit to the K07 data with $\chi^{2}/ \nu
=[1.62,1.44,0.90]$ at $\langle z \rangle = [2.07,2.52,2.94]$ for
$\nu=11$ degrees of freedom, corresponding to probabilities of
[9,15,54] per cent. These are therefore reasonable fits in a
quantitative sense.  We would,
however, urge the reader to exercise some caution in interpreting the
absolute $\chi^{2}$ values; estimating the observational PDF errors
and the covariance matrix is problematic.  We may have underestimated
the observational errors, especially at low flux levels.  Note that
the non-linear high density regions responsible for the absorption at
low flux levels are also the most sensitive to galactic winds and
other uncertain details of the numerical simulations.  It is
nevertheless gratifying that the best fitting values for
$\tau_{\rm eff}$ are all within 7 per cent of the mean values in each
redshift bin,  well within the variance of this quantity.  The model
with an almost isothermal temperature-density relation and $\sigma_8 =
0.95$ (20-256s2g2) is also in marginal agreement with the data.  Note
that although some of the remaining models are in reasonable agreement
with the observed PDF for the highest redshift bin $\langle z \rangle
= 2.94$, they are all in poor agreement if the full redshift range of
the K07 data set is taken into account.

In Table 4 we list the minimum reduced $\chi^{2}$ values for a
comparison of the simulated and observed spectra over the full range of the
PDF, $0\leq F \leq 1$.   The inverted temperature-density relation
again provides the best fit to the K07 data with $\chi^{2}/ \nu
=[3.06,1.56,1.82]$ at $\langle z \rangle = [2.07,2.52,2.94]$ for
$\nu=17$ degrees of freedom. The formal probabilities 
of the reduced  $\chi^{2}$ values are now rather low, with 
a very small probability at  $\langle z \rangle = 2.07$ and [7,2] per
cent probability at $\langle z \rangle = [2.52,2.94]$.  However,
as already discussed, estimating the errors of the PDF is difficult. 
Rather than being evidence against this model this  may 
suggest that we have somewhat underestimated the errors 
at the extreme ends of the PDF, especially at low flux levels. 
The best fitting values for $\tau_{\rm eff}$ are again within 6 per 
cent of the mean values in each redshift bin.  All the other models
are very bad fits.

\begin{figure*}
  \centering 
  \begin{minipage}{180mm} 
    \begin{center}

      \psfig{figure=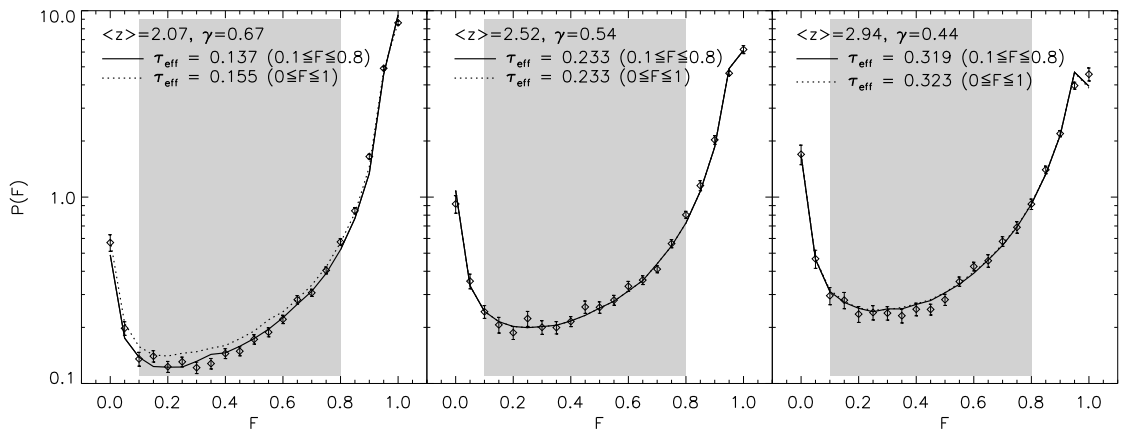,width=0.85\textwidth}
	    \caption{The best  fit to the PDF data of K07
	    (open diamonds with error bars) for the inverted
	    temperature-density relation model.  The solid curves show the best fits over
	    the flux range $0.1 \leq  F \leq 0.8$, corresponding to
	    the data in the shaded portion of the plots, while the
	    dotted curves show the best $\chi^{2}$ fits to the entire PDF. The
	    best fitting value of the effective optical depth for the
	    models are also displayed in each panel.}
      \label{fig:bestfit} \end{center} \end{minipage}
\end{figure*}
 
In  Fig.~\ref{fig:bestfit}, we show the best fits to the K07 flux 
PDF over the range $0.1\leq F \leq 0.8$ (solid curves) and $0\leq F
\leq 1$  (dotted curves).
The models are almost indistinguishable to the eye in the two highest
redshift bins, although one may clearly perceive the difference
between the different PDFs at $\langle z \rangle = 2.07$.  The poor
fit to the K07 data at $\langle z \rangle = 2.07$ for $0\leq F \leq 1$
is mainly due to the influence of the data point at $F=0$; the
agreement between the the observed and simulated PDF at $F>0.8$ is
very good.  Considering the results presented in
Fig.~\ref{fig:numerical}, a model with strong galactic winds may also
alleviate some of this discrepancy.
Even though we have attempted to accurately model
the noise properties of the observed spectra, differences in the noise
level in the spectra may also play a small role in this disagreement (K07).
In any case, the improved agreement between the observational data and 
the \Lya forest simulations with the inverted temperature-density
relation is impressive.

Finally, we note that the best fitting $\tau_{\rm eff}$ values for the
inverted temperature-density relation model correspond to a
photoionisation per hydrogen atom of $\Gamma_{\rm HI} =
[1.43,1.05,0.97] \times 10^{-12} \rm~s^{-1}$ for $0.1 \leq F \leq 0.8$
and $\Gamma_{\rm HI}=[1.14,1.05,0.95] \times 10^{-12} \rm~s^{-1}$ for
$0\leq F \leq 1$ at $\langle z \rangle = [2.07,2.52,2.94]$.  These
values are in excellent agreement with the results of \cite{Bolton05}
which are based on independent measurements of the \Lya forest opacity
(\citealt{Schaye03}).

\subsection{The effect of continuum uncertainties on the $\chi^2$ analysis}

We now consider how important the continuum placement on the observed
spectra is for the quality of the fit of the flux PDF at $0\leq F \leq
1$.  K07 discussed the effect of varying the continuum on the
observational data in detail.  For their typical estimated continuum
uncertainty of $\sim 1-2$ per cent, the changes in the PDF are small
and are most evident at $F>0.8$.  \cite{Tytler04} and \cite{Kirkman05}
also estimated continuum uncertainties to be of order $2$ per cent
based on an analysis of a large set of moderate resolution \Lya forest
spectra at $z<3.2$.  \cite{Faucher07} estimate the continuum placement
to be somewhat larger with a systematic bias towards an underestimate
of the continuum level by as much  $4$ per cent at $z=3$ and $1$ per
cent at $z=2$ in a recent measurement of the \Lya effective optical
depth using high-resolution data.  Note, however, that while a
systematic  misplacement of the continuum appears  likely  with
decreasing mean  flux level (and thus increasing redshift) it will 
also depend strongly on the shape of the \Lya flux PDF at high flux
levels and  thus on the temperature of the gas in low-density regions
of the IGM.

We have checked  for the presence of such a  potential bias   by
comparing all our  \Lya forest models to the K07 data with synthetic
spectra for which we have  raised or lowered the  continuum by 1.5 and
3 per cent.  Raising the continuum on the synthetic spectra
significantly worsens the agreement between the K07 data and all the
simulations.   Lowering the continuum on the synthetic spectra,
mimicking a continuum which has been placed too low on the
observational data, also strongly increases the $\chi^2$ values,
although the models with either an inverted or isothermal
temperature-density relation are still slightly preferred in some
redshift bins.  Misplacing the continuum changes the flux PDF in a
complex way.  The fact that the agreement between the observed data
and all models worsens significantly on varying the continuum suggests
that continuum fitting errors for the high-quality data of K07 are
under control at a level of 1-2 per cent, consistent with the
estimates of the continuum uncertainties by K07.  Altering the
continuum placement on the synthetic spectra does not alter our
conclusions regarding the best fitting model to the K07 data.
	
\subsection{Comparison to other studies} \label{sec:otherdata}

Several other detailed studies of the \Lya flux PDF are presented
within the existing literature, and we now discuss some of these
results in comparison to this work.  

\cite{McDonald00} (hereafter M00) presented a measurement  of the flux
PDF  at $\langle z \rangle =[2.41,3.00,3.89]$ from a sample of eight
high resolution QSO spectra.  A detailed comparison between the widely
used M00 measurement  and the measurement used in this work is given
in K07.  Here we comment on the interpretation of the M00 data when
compared to hydrodynamical \Lya forest simulations.  M00 compare their
observational data to the L10 Eulerian hydrodynamical simulation of
\cite{MiraldaEscude96}.  The simulation parameters were $\Omega_{\rm
m}=0.4$, $\Omega_{\Lambda}=0.6$, $\Omega_{\rm b}h^{2}=0.015$,
$h=0.65$, $\sigma_8=0.79$, $n=0.95$, $T_0=16~000 \rm~K$ and $\gamma
\simeq 1.3$ at $z =3$, with a box size of $10h^{-1}$ comoving Mpc and
a $288^{3}$ grid.  Performing a $\chi^{2}$ analysis with the
covariance matrix, M00 obtain $\chi^{2}/\nu=[4.4,1.7,1.1]$ for 19
degrees of freedom at $\langle z \rangle =[2.41,3.00,3.89]$.  Note,
however, that they do not minimise over $\tau_{\rm eff}$ as we do.
Instead, M00 rescale the synthetic spectra to match the measured
effective optical depth of the observational data, $\tau_{\rm
eff}=[0.201,0.380,0.744]$.  The fit is performed over the range $0\leq
F \leq 1$, and they attribute the disagreement at $\langle z \rangle =
2.41$ as being due to the effect of continuum uncertainties on the PDF.

As discussed in the last section this is in contrast with our
findings: using updated cosmological parameters we find the K07 PDF,
which is systematically lower at $0.2<F<0.7$ than the M00 data,
strongly favours an inverted temperature-density relation when fitting
over $0\leq F \leq 1$. It is not obvious if this difference  is due to
the different shape of the measured PDFs, differences between the
model parameters or differences in the numerical simulations.  To
investigate this further  we have attempted to fit the flux PDF of our
simulations  at $z=2.94$ to the M00 measurement of the flux PDF  at
$\langle z \rangle =3$.  The small difference in redshift should  not
be important here.  When minimising over $\tau_{\rm eff}$ we find that
all of the models listed in Table 3 are in good agreement with the M00
data for $0.1 \leq F \leq 0.8$ (all have $\chi^{2}/\nu \leq
1.04$). The inverted temperature-density relation model still provides
the best reduced $\chi^{2}$ ($\chi^{2}/\nu=0.41$ for $\tau_{\rm
eff}=0.411$) perhaps suggesting that the errors in this flux range
have been overestimated by M00.  When performing the $\chi^{2}$
analysis over the entire M00 PDF, we were at first not able to obtain
an acceptable fit to the M00 data.  Note, however, that M00 also
adjusted the normalisation of their synthetic spectra to take into
account possible continuum errors of the observational data.  They
achieved this by identifying the maximum transmitted flux in each
synthetic line-of-sight, $F_{\rm c}$, as the continuum flux and then
proceeded to divide the flux in all other pixels in the line-of-sight
by $F_{\rm c}$.

We therefore again  raise or lower the continuum in our synthetic
spectra by 1.5 per cent, this time for comparison to the M00 data at
$\langle z \rangle = 3$.  As for the K07 data, raising the continuum
on the synthetic spectra significantly worsens the agreement between
the M00 data and all our  models.  Lowering the continuum in this
instance, however, improves the agreement between the M00 data and
simulations remarkably.  Assuming our synthetic spectra provide an
accurate representation of the \Lya forest, this suggests that the
continuum in the absorption spectra on which the M00 measurement is
based may indeed be placed somewhat low.  If we lower the continuum
level by 1.5 per cent, we find the model with a temperature-density
relation close to isothermal  and a larger $\sigma_8=0.95$
(20-256s2g2) provides the best fit to the M00 data, with
$\chi^{2}/\nu=1.08$ for an effective optical depth $\tau_{\rm
eff}=0.385$, which is pleasingly close to the M00 value.  The inverted
temperature-density relation model (20-256g3) provides the next best
fit with $\chi^{2}/\nu=1.91$ for $\tau_{\rm eff}=0.418$.  This is in
contrast to M00, who achieved a reasonable fit to the data using
different (now somewhat outdated)  cosmological parameters and a
standard temperature-density relation. The different conclusion we
reach in this work is therefore probably largely due to the updated
cosmological parameters and perhaps also numerical differences in the
simulations.  In summary, we conclude that the M00 PDF measurements
are consistent with our numerical simulations if the continuum is
lowered by $1.5$ per cent. The M00 measurement  also appears to favour
a flattened or inverted temperature-density relation at $z \simeq 3$,
consistent with our results based on the independent K07 data set.

A more recent joint analysis of the best fitting model to the M00 PDF
and flux power spectrum was undertaken by \cite{Desjacques05} at
$z=3$.  They used N-body simulations with $256^{3}$ dark matter
particles and a box size of $25h^{-1}$ comoving Mpc.  Synthetic \Lya
forest spectra were constructed by smoothing the dark matter density
and velocity fields with a Gaussian filter over a characteristic
scale, which was left as an extra free parameter in their
analysis. Unfortunately,  this extra free parameter complicates our
comparison to this work as it is degenerate with other relevant
quantities, most notably $\gamma$.  Consequently, \cite{Desjacques05}
do not advocate a value for this quantity.  On the other hand,
\cite{Desjacques05} also found that their joint analysis of the M00
flux power spectrum and PDF (fitted over the range $0.05 \leq F \leq
0.75$) may favour a lower value of $\sigma_8$ than previous analyses of
the power spectrum alone (\citealt{Viel04,McDonald05}), although this
result should  again  be degenerate with the  thermal state of the
IGM.  We instead find that the K07 PDF measurements on their own favour
value of $\sigma_8$ towards the upper end of current observational
constraints if the temperature-density relation is close to isothermal.

\cite{Lidz06b} compare synthetic spectra drawn from
pseudo-hydrodynamical simulations to the M00 observational data, but
instead use a different estimator for the flux PDF.  This estimator is
derived from data which is smoothed on two different scales, thereby
having the advantage of rendering the observational data insensitive
to the normalisation {\it and} shape of the continuum, but also to structure
on small scales in the spectra.  This latter effect may be
advantageous when using simulations which are not fully resolved, but
it may also erase important information in the observational data,
especially with regard to the thermal state of the IGM.
\cite{Lidz06b}  find their simulations favour a temperature-density
relation with $\gamma\geq 1.32$ ($2\sigma$) at $z=2.72$.  However,
given the approximate treatment of smoothing due to gas pressure
\cite{Lidz06b} adopt in their simulations, as well as their different
observational estimator for the PDF, the significance of this
constraint on comparison to our results is difficult to judge.

Lastly, and as discussed previously, \cite{Becker07} found that an
inverted temperature-density relation may be required to adequately fit
the flux PDFs measured from an independent set of high resolution QSO
spectra over a wide redshift range, $1.7<z<5.8$. Our results are also
consistent with this possibility. Note, however, that \cite{Becker07}
use a semi-analytical model for the \Lya forest based on the IGM
density distributions drawn from a hydrodynamical simulation.  This
model does not self-consistently model Jeans smoothing effects, and a
direct comparison to their results is again not possible.

\section{Implications for the \Lya forest flux power spectrum}

The \Lya flux PDF is of course not the only diagnostic of the  thermal
state of the IGM. There should be good prospects to break  some of the
degeneracies discussed here and further tighten constraints by a
combined analysis with other \Lya flux statistics.  Assumptions about
the IGM thermal state are of particular relevance for measurements of
the matter power spectrum from the \Lya flux power spectrum. Such
analyses have so far not allowed for the possibility of an inverted
temperature-density relation and have marginalised over what was
believed to be the plausible range for the slope, $1<\gamma<1.6$. Due
to the temperature dependence of the recombination coefficient a
smaller $\gamma$ should translate into a steeper power law relation
between neutral hydrogen density and matter density. For measurements
of the matter power spectrum a smaller assumed $\gamma$ should
therefore result in a smaller amplitude for the matter power spectrum.
In this section we therefore briefly discuss the implications of an
inverted temperature-density relation for the \Lya forest flux power
spectrum.   Note this is a qualitative comparison only; a detailed
joint analysis of the \Lya flux PDF and flux power spectrum with a
larger suite of numerical simulations with inverted
temperature-density relations will be required to examine this more
accurately, which is beyond the scope of this paper.

\begin{figure}
\begin{center} 
  \includegraphics[width=0.45\textwidth]{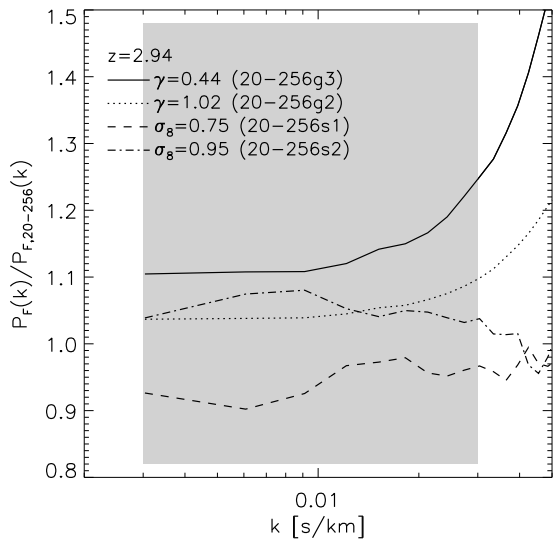} \caption{The ratio
  of the \Lya flux power spectrum computed from synthetic spectra
  drawn from the models indicated on the panel at $z=2.94$ to the flux
  power spectrum of the fiducial model (20-256, $\sigma_8=0.85$,
  $\gamma = 1.27$).  The shaded region corresponds to the
  range of wavenumbers used by Viel et al. (2004b) to infer the
  amplitude and shape of the matter power spectrum from the \Lya flux
  power spectrum, $0.003<k({\rm s~km^{-1}})<0.03$.  An inverted
  temperature-density relation increases the power within this
  wavenumber interval, mimicking the effect of a larger amplitude
  $\sigma_8$ of the underlying matter power spectrum.}
\label{fig:power}
\end{center} 
\end{figure}

Fig.~\ref{fig:power} illustrates the effect of an inverted
temperature-density relation on the flux power spectrum at $z=2.94$.
The solid curve shows the ratio of the flux power spectrum, $P_{\rm
F}(k)$, computed from the model with the inverted temperature-density
relation to the fiducial model power spectrum, $P_{\rm F,20-256}(k)$.
The shaded region corresponds to the range of wavenumbers used by
\cite{Viel04b} to infer the matter power spectrum from the \Lya forest
flux power spectrum. The flux power spectrum of the simulations with
the inverted temperature-density relation clearly
exhibits more power over this range on comparison to the fiducial
model, and this increases noticeably towards larger
wavenumbers.  As discussed in section~\ref{sec:params}, this is
because a larger $\gamma$ increases the temperature and hence
the ionised fraction of the gas associated with the most prominent
absorption features, thereby suppressing predominantly the small scale
power.  The dotted curve shows the same comparison for the isothermal
temperature-density relation model.  The effect is again to increase
power at all scales, although to a somewhat lesser extent.  The dashed
and dot-dashed curves also show the comparison for two models with
different matter power spectrum normalisations, $\sigma_8=0.75$ and
$\sigma_8=0.95$, respectively.  As expected, a lower $\sigma_8$
suppresses power and a larger $\sigma_8$ increases power over the
relevant wavenumber range.   An inverted or isothermal
temperature-density relation therefore mimics some of the effects a
larger normalisation for the matter power spectrum has on the \Lya forest
flux power spectrum.

This has potentially important implications for inferring the matter
power spectrum from the \Lya forest.   The measurements of the matter
power spectrum from \Lya forest data are affected by non-linearities
as well as by complex degeneracies between the thermal state of the
IGM, the density distribution and the assumed effective optical depth.
However, if we naively extrapolate the results of \cite{Viel04b} to
smaller values of $\gamma$ we would expect a 5-10 per cent percent
lower amplitude $\sigma_8$ for the slope of $\gamma \sim 0.5$ favoured
by the analysis of the \Lya flux PDF presented here.  A 5-10 per cent
lower amplitude for the inferred matter power spectrum  would largely
alleviate the strain between the measurements from \Lya forest data
and the WMAP 3 year results (\citealt{VielHaehnelt06}).  Unfortunately
it  would also substantially weaken the strong upper limits on the
neutrino mass from the joint analysis of \Lya forest and CMB data
(\citealt{Seljak06b,Gratton07}).

\section{Conclusions}

We have compared the improved measurements of the \Lya flux PDF at
$1.7<z<3.2$ by K07 obtained from a full Voigt profile analysis  of 18
high resolution, high signal to noise  UVES/VLT spectra to a large set
of hydrodynamical simulations of the IGM.  Our main findings are as
follows:

\begin{itemize} 

\item {The shape of the \Lya flux PDF depends only weakly on the
fluctuation amplitude of the matter power spectrum characterised by
$\sigma_{8}$ and on the normalisation of  the temperature-density
relation for the low density IGM, $T=T_0 \Delta^{\gamma-1}$
(\citealt{HuiGnedin97,Valageas02}).  A moderately increased  scatter
in the temperature-density relation, which may result from
inhomogeneous radiative transfer effects during the epoch of \HeII
reionisation, also has little effect on the shape of the flux PDF.
The shape of the \Lya flux PDF is mainly determined by the  effective
optical depth $\tau_{\rm eff}$ and by the slope of the
temperature-density relation for the low density IGM.  The extreme
ends of the PDF at flux levels of zero and unity are particularly
sensitive to the slope of the temperature-density relation.\\}
 
\item{A comparison of the full range of the \Lya flux PDF to
hydrodynamical simulations favours an inverted temperature-density
relation.     The flux PDFs of our synthetic
spectra with an inverted temperature-density relation, such that $\gamma =
[0.67,0.54,0.44]$ at $\langle z \rangle =[2.07,2.52,2.94]$,  provide
an excellent fit to the observed metal-cleaned flux PDF measured by
K07.   This suggests that the voids in the IGM  are significantly
hotter than  previously assumed.\\}

\item {If we restrict our analysis of the PDF to flux levels  $0.1
\leq F \leq 0.8$,  which are least affected by continuum fitting,
noise, galactic winds  and the uncertainties of how to simulate the
relevant physics in dense regions of the IGM, the comparison of
observed  and synthetic spectra  still  favours a model where the
temperature-density relation is either nearly isothermal or inverted.\\}

\item{Raising or lowering the assumed continuum level of our synthetic
spectra in order to mimic the possibility of a systematic over or
underestimate  of the continuum level in the observed spectra rapidly
worsens the quality of the fit to the K07 PDF for offsets larger than
1.5 per cent. This suggests that continuum fitting errors for the
high quality data of K07 are under control at a level of 1-2 per cent,
consistent with the estimates of the continuum uncertainties by K07.\\}

\item{A comparison of our simulations with the full range of the  \Lya
flux PDF measured by M00 at $\langle z \rangle = 3$ from a smaller
observational sample gives consistent results if the continuum on the
synthetic spectra is lowered by $1.5$ per cent.  Using the currently
preferred cosmological parameters, the M00 data also favour a nearly
isothermal or inverted temperature-density relation.\\}

\item{An inverted temperature-density relation has important
implications when inferring the underlying matter power spectrum from
the \Lya forest flux distribution.  The effect of an inverted
temperature-density relation on the flux power spectrum is similar to
the effect of an larger normalisation, $\sigma_8$, of the matter power
spectrum.  This  could alleviate some of the existing strain between
the measurements of $\sigma_8$ from \Lya forest data and the WMAP 3
year results  (\citealt{VielHaehnelt06}).}
     
\end{itemize}

The preference of the data for an inverted temperature-density
relation would, if consolidated, be a rather profound result.
It would suggest that the widely accepted paradigm of a tight IGM
temperature-density relation at low densities, evolving from being
almost isothermal following hydrogen reionisation to a relation where
the temperature rises  with increasing density
(\citealt{HuiGnedin97,Theuns98,Valageas02}), is too simplistic.

Perhaps the most plausible explanation for an inverted temperature
density relation is \HeII reionisation. There is evidence to suggest
this may end around $z\sim 3$
(\citealt{Songaila98,Schaye00,Theuns02d,Shull04,Bolton06,Faucher07}), and thus
lies at the upper end of the redshift range considered in this work.
Changes in the photoheating rate which occur as ionising photons
propagate through optically thick regions can significantly raise the
temperature of the IGM (\citealt{AbelHaehnelt99}). First attempts at
modeling inhomogeneous \HeII reionisation and/or the associated
radiative transfer effects also suggest a more complex and possibly
multi-valued and inverted temperature-density relation is plausible
(\citealt{Bolton04,Gleser05,Tittley07, Furlanetto07}). Therefore,
while a well-defined, inverted temperature-density relation may not
faithfully represent the true thermal state of the IGM, a model with
$\gamma<1$ should mimic some of the expected  inhomogeneous radiative
transfer induced effects on the IGM temperature following \HeII
reionisation.  Within this framework, this model provides a good fit
to the K07 observational data and suggests that the thermal  state of
the IGM may be significantly more complex at $z\simeq 3$ than
previously thought.    Further investigation of the thermal state of
the IGM around $z\simeq 3$ is therefore highly desirable. Note,
however,  that  studies using wavelets have failed to  detect the
expected fluctuations in the IGM  temperature at these  redshifts
(\citealt{Theuns02c,Zaldarriaga02}).

While radiative transfer effects and the inhomogeneity of \HeII
reionisation provide a plausible explanation for an increased
temperature in underdense regions  of the IGM, it will  require full
numerical radiative transfer simulations to confirm the validity of
this picture. Helium reionisation is driven by hard photons emitted by
rare, luminous and most likely short-lived sources (QSOs) under
conditions where photon mean free paths are significantly longer and
recombination time scales shorter (if compared to the Hubble time)
than for hydrogen reionisation at  higher redshifts. Current 3D
cosmological radiative transfer codes which incorporate multiple
sources are designed primarily to investigate the topology of \HI
reionisation, and are less reliable when modeling the resulting
thermal state of the IGM (\citealt{Iliev06}).  Most detailed studies
of radiative transfer effects on the IGM temperature therefore employ
either reduced dimensionality (\citealt{Bolton04,BoltonHaehnelt07}) or
neglect the modeling of multiple sources
(\citealt{Maselli05,Tittley07}). Furthermore, the physical scales
involved during \HeII reionisation are much larger than those during
\HI reionisation at $z>6$, placing even greater demands on the
required computational resources (although see \citealt{Sokasian02,Paschos07}).
Radiative transfer simulations which accurately  track the
temperature evolution  are thus very challenging.   Developments in
this area will be  key to interpreting the increasing amount of high
quality \Lya forest data.

Finally, we must also not lose sight of the possibility that other
processes may contribute to the heating of the IGM (e.g.
\citealt{Nath99,MadauEfstathiou99,Inoue03,RicottiOstriker04,Samui05}).
Our results may instead indicate that an alternative modification to
the current models for the \Lya forest and the thermal state of the
IGM is required.

\section*{Acknowledgements}

We thank Volker Springel for his advice and for making $\rm
\scriptstyle GADGET-2$ available, and Benedetta Ciardi and Simon White
for comments on the draft manuscript.  We also thank the anonymous
referee for a positive and helpful report.  This research was
conducted in cooperation with SGI/Intel utilising the Altix 3700
supercomputer COSMOS at the Department of Applied Mathematics and
Theoretical Physics in Cambridge.  COSMOS is a UK-CCC facility which
is supported by HEFCE and STFC/PPARC.


\end{document}